\documentclass[conference]{IEEEtran}
\IEEEoverridecommandlockouts
\usepackage{amsmath,amsfonts,amsxtra,amssymb,latexsym,amscd,amsthm,mathrsfs,bm}
\DeclareUnicodeCharacter{2212}{-}
\usepackage{graphicx}
\usepackage{xcolor}
\usepackage{cite}
\usepackage{cuted}
\usepackage[small]{caption}
\usepackage{algorithm}
\usepackage{multirow}
\usepackage{comment} 
\usepackage{lipsum}
\usepackage{optidef}
\usepackage{mathtools}
\usepackage{multicol}

\usepackage{algpseudocode}
\usepackage{verbatim}

\def\BibTeX{{\rm B\kern-.05em{\sc i\kern-.025em b}\kern-.08em
    T\kern-.1667em\lower.7ex\hbox{E}\kern-.125emX}}
    
\begin{document}

\title{{Low Complexity Lookup Table Aided Soft Output Semidefinite Relaxation based Faster-than-Nyquist Signaling Detector}\\
\thanks{This work was funded by the Scientific and Technological Research Council of Turkey (TUBITAK) Project Grant No. 122E236 and TUBITAK 2214/A International Doctoral Research Fellowship Programme.}}

\author{
\IEEEauthorblockN{Adem Cicek}
\IEEEauthorblockA{\textit{Electrical and Electronics Eng. Dept.} \\
\textit{Ankara Yildirim Beyazit University}\\
Ankara, Turkey \\
ademcicek@karatekin.edu.tr}
\and
\IEEEauthorblockN{Ian Marsland}
\IEEEauthorblockA{\textit{Systems and Computer Eng. Dept.} \\
\textit{Carleton University}\\
Ottawa, ON, Canada \\
ianm@sce.carleton.ca}
\and
\IEEEauthorblockN{Enver Cavus}
\IEEEauthorblockA{\textit{Electrical and Electronics Eng. Dept.} \\
\textit{Ankara Yildirim Beyazit University}\\
Ankara, Turkey \\
ecavus@aybu.edu.tr}
\and
\IEEEauthorblockN{\hspace*{2.0cm}Ebrahim Bedeer}
\IEEEauthorblockA{\textit{\hspace*{2.0cm}Electrical and Computer Eng. Dept.} \\
\hspace*{2.0cm}\textit{University of Saskatchewan}\\
\hspace*{2.0cm}Saskatoon, SK, Canada \\
\hspace*{2.0cm}e.bedeer@usask.ca}
\and
\IEEEauthorblockN{\hspace*{-2cm}Halim Yanikomeroglu}
\IEEEauthorblockA{\hspace*{-2cm}\textit{Systems and Computer Eng. Dept.} \\
\hspace*{-2cm}\textit{Carleton University}\\
\hspace*{-2cm}Ottawa, ON, Canada \\
\hspace*{-2cm}halim@sce.carleton.ca}
}
\maketitle
\begin{abstract}
Spectrum scarcity necessitates innovative, spectral-efficient strategies to meet the ever-growing demand for high data rates. Faster-than-Nyquist (FTN) signaling emerges as a compelling spectral-efficient transmission method that pushes transmit data symbols beyond the Nyquist limit, offering enhanced spectral efficiency (SE). While FTN signaling maintains SE with the same energy and bandwidth as the Nyquist signaling, it introduces increased complexity, particularly at higher modulation levels. This complexity predominantly arises from the detection process, which seeks to mitigate the intentional intersymbol interference generated by FTN signaling. Another challenge involves the generation of reliable log-likelihood ratios (LLRs) vital for soft channel decoders. In this study, we introduce a lookup table (LUT) aided soft output semidefinite relaxation (soSDR) based sub-optimal FTN detector, which can be extended to higher modulation levels. This detector possesses polynomial computational complexity, given the negligible complexity associated with soft value generation. Our study assesses the performance of this soft output detector against that of the optimal FTN detector, Bahl, Cocke, Jelinek and Raviv (BCJR) algorithm as the benchmark. The likelihood values produced by our LUT aided semidefinite relaxation (SDR) based FTN signaling detector show promising viability in coded scenario. 
\end{abstract}
\section{Introduction}
With the escalating demands for higher data rates and finite spectrum resources, there is an imperative to develop bandwidth-efficient transmission techniques. Faster-than-Nyquist (FTN) signaling emerges as a potent solution in this space, using non-orthogonal pulses in the time domain to transmit more data within the same bandwidth and signal-to-noise ratio (SNR) \cite{anderson2013faster}. Recent attention from both industry and academia has revived interest in FTN signaling, spotlighting its potential in various applications such as point-to-point microwave backhaul links, optical long-haul links, digital video broadcasting, satellite communications, and next-generation cellular networks with massive antenna arrays at base stations\cite{modenini2014faster, ZouOpticFTN}. While FTN signaling enhances the SE, it introduces controlled inter-symbol interference (ISI) between neighboring pulses at the receiver {\cite{anderson2013faster}, \cite{CicekCIFTN}, \cite{David8805289}}. Addressing this ISI necessitates advanced signal processing techniques, which in turn add to the detection complexity. At the same time, a significant advancement towards the practical deployment of FTN signaling is the development of FTN signaling detectors that produce soft output decisions. But generating soft outputs also introduces its own set of challenges, especially in crafting log-likelihood ratio (LLR) values for channel decoders in coded setups. The precise LLR calculation for a single bit in an $N$-bit sequence involves processing $2^N$ hard-output sequences, meaning a daunting computational task. In the FTN signaling context, while the coded studies that employs the BCJR algorithm as the FTN detector have already reliable LLR values, the coded studies that propose new low-complexity detectors finds the LLR values by reducing the search space not to increase more complexity. \cite{Sina_DL_based_LSD} that uses the list-sphere decoding as a detector finds the LLR values after pre-determining initial distance to give $N_L$ sequences with the help of deep learning, and if the decoder can not obtain $N_L$ sequences in the predetermined distance it increases the search distance to get enough sequences, yielding to $\mathcal{O}(N_L N^2)$ compared to $\mathcal{O}(2^N N^2)$ required for the calculation of the exact LLR values. The work in \cite{Sina_classification_approach} that proposes the classification approach and classifies $N_p$-dimensional sequence instead of $N$-dimensional to detect each symbol taking the $N_L$ sequences into account to calculate the soft values, necessitating the complexity of $\mathcal{O}(N_L N_p^2)$. LLR generation studies for BPSK modulation will be much more complex for higher modulation.\\
\indent This paper introduces a low-complexity lookup table (LUT) aided soft-output semidefinite relaxation (soSDR) based detector, tailored for coded FTN signaling systems. SDR-based FTN detectors, as highlighted in \cite{bedeer2017low} and \cite{bedeer2019low}, are commendable due to their good performance combined with polynomial time complexity. Our proposed soft-output generation method that fully works offline furnishes dependable LLR values for the channel decoder. This is achieved without inflating the complexity compared to the innate complexity of the semidefinite relaxation optimization. To validate the efficiency of our proposed method, we simulated the soSDR FTN signaling detector within the context of polar codes for channel coding, comparing its performance against the benchmark BCJR detector. Simulations indicate that the soSDR detector generates reliable soft values, striking a good balance between bit error rate (BER) performance and detection complexity.\\
\indent The structure of this paper is as follows: Section \ref{sec:modell} reviews the coded FTN signaling system model. The proposed soft-output semidefinite relaxation-based detector is elaborated in Section \ref{sec:sosdr}. Section \ref{sec:complexity_sdr} presents the complexity comparison of the soSDR against the BCJR, followed by simulation outcomes in Section \ref{sec:simulation}. The paper is concluded in Section \ref{sec:conc}.\\
\indent The following notation is used throughout the paper: The operators $\bf{tr}(\cdot)$, $||\cdot||$, $Pr(\cdot)$, $p(\cdot)$ and ${(\cdot)}^{H}$ represent the trace of a matrix, the Euclidean norm, the probability, the probability density function and the Hermitian transpose, respectively. Uppercase bold letters denote matrices whereas column vectors are shown by lowercase bold letters. Scalar variables are indicated in italic font.
\section{System Model}\label{sec:modell}

\begin{figure}[t]
   \centering
    \includegraphics[width=0.45\textwidth]{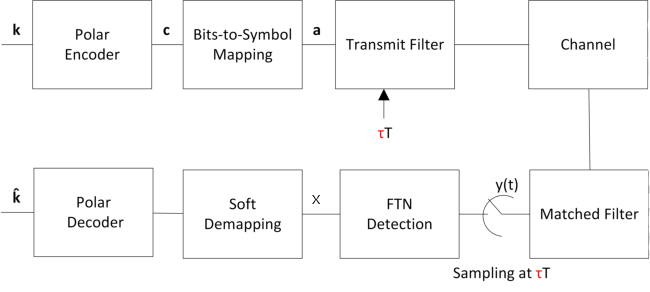}
   \caption{Block diagram of channel coded FTN signaling.}
    \label{system_modell}
\end{figure}

Fig. \ref{system_modell} depicts a block diagram of a polar coded communication system using FTN signaling. On the transmitter side, a sequence \(\bf{k}\) comprising \(K\) information bits undergoes polarization through a polar encoder, leveraging a recursive butterfly structure. This process polarizes \(N\) channels into highly reliable and completely unreliable channels, respectively. The encoder yields \(N\) bits, allocating the \(K\) most reliable bit positions to information bits, termed as the free bits. The remaining \(N-K\) positions, recognized as frozen bits, are set to a value of 0. The bits placed onto a vector \(\bf{u}\) are encoded as
\begin{equation}\label{encoding}
    \bf{c} = \bf{u} \bf{G},
\end{equation}
where $\bf{G}$ represents the polar code generator matrix.

Each bit in the codeword \(\bf{c}\) undergoes $M$-ary pulse amplitude modulation ($M$-PAM), followed by shaping via a unit energy pulse \(z(t)\). Please note that all of the equations are valid for other $M$ values, and can also be extended to $M$-ary quadrature amplitude modulation ($M$-QAM), with some modifications. $M=4$ is chosen to make it easier to understand with channel coding. Then, symbols are transmitted every \(\tau T\), where \(0 < \tau \leq 1\) signifies the FTN signaling acceleration parameter, and \(T\) is the symbol duration. After this process, the signal travels through an additive white Gaussian noise channel (AWGN), represented as \(w(t)\).
At the receiver, to retrieve the transmitted symbols from the noisy signal, a matched filter is employed. The output from this filter is
\begin{equation}\label{matched_filt_output}
    y(t) = \sqrt{\frac{K}{N} E_{s}} \sum_{n=1}^{N} a_{n} g(t-n\tau T) + q(t),
\end{equation}
where $\bf{a}$ is the sequence of symbols whose elements are the $4$-PAM data, $a_{n}$, $n = 1,2,..,N$, and $E_{s}$ is the energy of the data symbol, $\frac{K}{N}$ is the code rate, $g(t)$ and $q(t)$ are the convolution of $z(t)$ and $w(t)$, respectively, with the matched filter.
The signaling rate is \( \frac{1}{\tau T} \). The received signal, denoted by \(y(t)\), undergoes sampling every \(\tau T\) seconds, and the \(l\)th received sample can be described by
\begin{equation}\label{received_sample}
\begin{aligned}
y_{l} &=  y(l\tau T) \\ &= \sqrt{\frac{K}{N} \: E_{s}}\sum_{n=1}^{N}a_{n} g(l\tau T-n\tau T) + q(l\tau T) \\
& = \underbrace{\sqrt{\frac{K}{N} \: E_{s}}a_{l}g(0)}_{\text{desired symbol}} \\
& + \underbrace{\sqrt{\frac{K}{N} \: E_{s}}\sum_{n=1, n\neq l}^{N}a_{n} g(l\tau T-n\tau T)}_{\text{ISI}} \\
& + \underbrace{q(l\tau T).}_{\text{ $l$th noise sample after matched filter}}
\end{aligned}
\end{equation}
From (\ref{received_sample}), the received signal encompasses the desired symbol, noise, and the Inter-Symbol Interference (ISI). The signal can be vectorized as
\begin{equation}\label{vector_form}
    {\bf{y}} = \sqrt{\frac{K}{N} E_{s}} {\bf{a}} * {\bf{g}}  + {\bf{q_{c}}},
\end{equation}
where $\bf{q_{c}}$ is the colored noise, * is the convolution operator, and $\bf{g}$ is the ISI vector.
Subsequently, we can represent (\ref{vector_form}) in the form
\begin{equation}\label{SMF_output}
    \bf{y} = \bf{G} \bf{a} + \bf{q_c},
\end{equation}
where $\bf{G}$ is the ISI matrix and $\bf{q_{c}}$ is the colored noise with probability density function of $\mathcal{N}\bf{(0, \sigma^2 \bf{G})}$.
After an approximate whitening matched filter\cite{PrljaWMF} is used the received vector becomes 
\begin{equation}\label{WMF_output}
    \bf{y} = \bf{V} \bf{a} + \bf{q_w},
\end{equation}
where $\bf{V}$ denotes the newly constructed ISI matrix based on the causal ISI vector. $\bf{q_{w}}$ represents the whitened noise, having a probability density function (pdf) $\mathcal{N}\bf{(0, \sigma^2I )}$, with $\textbf{I}$ being the identity matrix. This allows us to formally present the maximum likelihood sequence estimation (MLSE) problem of detecting $M$-PAM FTN signaling at the receiver as
\begin{equation}
\label{mlse_problem}
\begin{aligned}
& \underset{\bf{a} }{\text{minimize}}
& & \|\bf{y}-\bf{V} \bf{a}\|_{\text{2}}^{\text{2}},
\end{aligned}
\end{equation}
where {\bf{a}} represents an $N \times 1$ symbol vector derived from the relevant alphabet constellation. To address the detection problem presented in (\ref{mlse_problem}), we propose an SDR-based FTN signaling detector. This detector efficiently produces soft outputs utilizing lookup tables (LUTs), ensuring a minimal increase in computational complexity compared to its hard-output counterpart.

\begin{figure}[h]
\centering
\includegraphics[width=0.45\textwidth]{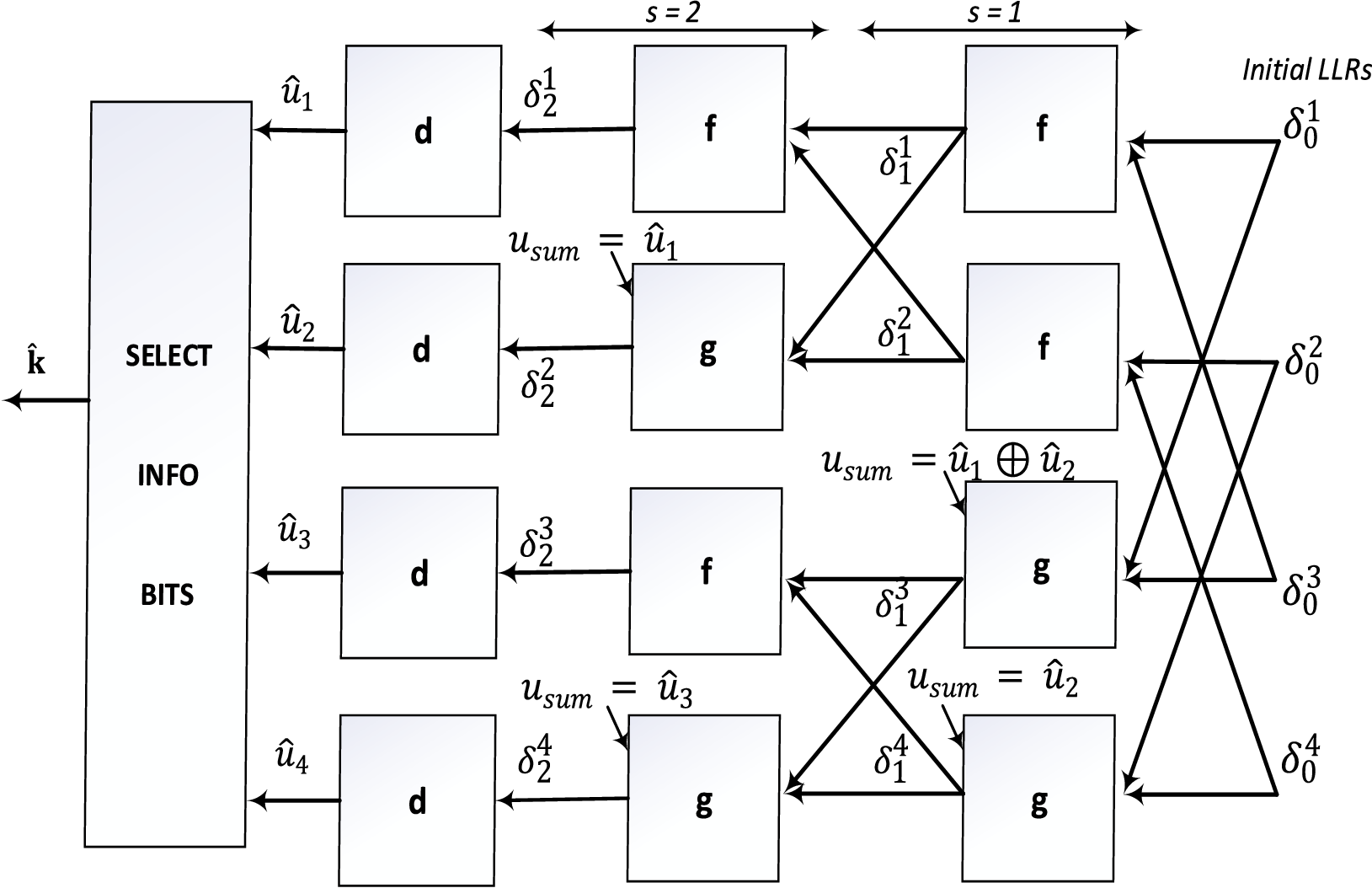}
\caption{Successive cancellation decoding of polar codes for $N$ = 4 bits.}
\label{Successive cancellation decoding}
\end{figure}
Soft outputs derived from solving the detection problem in (\ref{mlse_problem}) serve as the initial LLR input to the polar decoder. Within this polar decoder, the successive cancellation decoding algorithm (SCD) is employed to estimate the bit sequence $\bf{\hat{k}}$ \cite{leroux2011hardware}. Fig. \ref{Successive cancellation decoding} provides an illustration of a $4$-bit SCD example, utilizing three distinct functions: $f$, $\textsl{g}$, and $d$. Inputs $\delta^i_{0}$ and $\hat{u}_i$ ($i = 1,2,3,4$) correspond to the initial LLR value and the hard decision for the $i$th bit, respectively. 
Functions $f$ and $g$ essentially reverse the encoder operations as detailed in (\ref{llr_f_g}) to (\ref{g_func}). These operations use the LLR values from the preceding stage and the partial sums of earlier decoded bits ($u_{\rm{sum}}$) to determine the LLR values for the subsequent stage
\begin{equation}\label{llr_f_g}
\begin{aligned}
{\rm{\delta}}^{i}_{j+1} =
\begin{cases}
f\left({\rm{\delta}}^{i}_{j}, {\rm{\delta}}^{i+\frac{N}{2^{j+1}}}_{j}\right),\\
\textsl{g}\left({\rm{\delta}}^{i-\frac{N}{2^{j+1}}}_{j},{\rm{\delta}}^{i}_{j}, \hat{u}_{\rm{sum}}\right),
\end{cases}
\end{aligned}
\end{equation}
\begin{equation}\label{f_func}
\begin{aligned}
f(\mu,\nu) = {\rm{sign}}(\mu) \: {\rm{sign}}(\nu)\: {\min}(|\mu|, |\nu|),
\end{aligned}
\end{equation}
\begin{equation}\label{g_func}
\begin{aligned}
\textsl{g}(\mu,\nu,u_{\rm{sum}}) = (-1)^{u_{\rm{sum}}}\mu + \nu,  
\end{aligned}
\end{equation}
where ${\rm{\delta}}^{i}_{j+1}$ is the LLR value of $i$th bit at $(j+1)$th stage. $\rm{sign}(\cdot)$ gives the sign of ($\cdot$) while $\min(\cdot,\cdot)$ produces the minimum. In (\ref{g_func}), the partial sum of the previously decoded bit, $u_{\rm{sum}}$, is propagated back into the LLR estimation of the next stage. 
The $u_{\rm{sum}}$ values of stage 1 and 2 are given as
\begin{equation}\label{u_sum}
\begin{aligned}
{u}_{sum} = 
\begin{cases}
& \hat{u}_{1}\oplus\hat{u}_{2} \text{ , } \hat{u}_{2}\text{ at } s = 1,\\
& \hat{u}_{1} \text{ , } \hat{u}_{3}\text{ at } s=2,
\end{cases}
\end{aligned}
\end{equation}
where $\oplus$ is a logical xor operation. 
In this way, all bits are decoded at $s = \log_2N$ stages successively. Finally, the hard decisions $\hat{u}_{1}^{N}$, $i = 1, 2, ..., N-1, N$ based on the updated LLR values $\delta_{j+1}^{i}$, $i = 1, 2, ..., N; j = 0, 1, ..., s-2, s-1$ are calculated as in (\ref{d_func}) and then the information sequence $\bf{\hat{k}}$ (free bits) is selected from $\hat{u}_{1}^{N}$ according to their indices.
In the process of decoding, each bit is successively decoded over $s$ stages. Finally, based on the updated LLR values, the hard decisions are derived as demonstrated in (\ref{d_func}). Subsequently, the information sequence $\bf{\hat{k}}$ is extracted from $\hat{u}_{1}^{N}$ based on their indices:
\begin{equation}\label{d_func}
\begin{aligned}
\hat{u}_{i} = 
\begin{cases}
& 1, \delta_s^{i}\geq 0 \\
& 0, \delta_s^{i} < 0. 
\end{cases}
\end{aligned}
\end{equation}
\section{Low-complexity soft-output SDR based FTN detector}{\label{sec:sosdr}}
In this section, we present a soft-output SDR-based FTN detector, which has a similar polynomial detection complexity order to the hard-output SDR detector. The objective function described in (\ref{mlse_problem}) can be reformulated as 
\begin{equation}\label{new_obj_func}
\begin{aligned}
\bf{\left\| y- Va \right\|^2} &= (\bf{y-Va})^{\textit{H}} (\bf{y-Va})\\
&= \bf{y}^{\textit{H}}\bf{y} - \bf{y^{\textit{H}}Va} -\bf{a^{\textit{H}}V^{\textit{H}}y} 
+ \bf{a^{\textit{H}}V^{\textit{H}}Va}\\
&= \bf{\left\|y\right\|^2} - \bf{2a^{\textit{H}}V^{\textit{H}}y} + \bf{tr\lbrace V^{\textit{H}}VA\rbrace},
\end{aligned}
\end{equation}
where $\bf{A} = \bf{aa^{\textit{H}}}$ is $N \times N$ symmetric positive semidefinite matrix. Since $\bf{\left\|y\right\|^{2}}$ is independent of $\bf{a}$, the $M$-PAM FTN signaling detection problem can be expressed as
\begin{equation}\label{MLSE_to_detect_FTN}
\begin{aligned}
& \hat{\bf{a}} = \text{arg} \underset{\bf{a} \in {\cal{M}}^{\textit{N}\text{ x }\text{1}}} {\text{min}}
& & \bf{tr\lbrace V^{\textit{H}}VA\rbrace} - \bf{2a^{\textit{H}}V^{\textit{H}}y},
\end{aligned}
\end{equation}
where $\cal{M}$ represents the modulation alphabets. The FTN signaling detection problem in (\ref{MLSE_to_detect_FTN}) is a nonconvex problem with rank one constraint due to $\bf{A} = \bf{aa^{\textit{H}}}$; hence, it is hard to solve. We relax this rank one constraint, \cite{bedeer2019low} adding a convex positive semidefinite constraint of $\bf{A} - \bf{aa^{\textit{H}}}$ $\succeq$ $0$. Now, the general optimization form of the FTN signaling detection problem in (\ref{MLSE_to_detect_FTN}) can be written as
\begin{equation}\label{MLSE_under_SDR}
\begin{aligned}
& \hat{\bf{a}} = \text{arg} \underset{\bf{a} \in {\cal{M}}^{\textit{N}\text{ x }\text{1}}} {\text{min}} 
 \bf{tr\lbrace V^{\textit{H}}VA\rbrace} - \bf{2a^{\textit{H}}V^{\textit{H}}y} \\
& \text{subject to } \textit{constraints of the modulation schemes}\\
& \quad \quad \quad \quad \bf{A} - \bf{aa^{\textit{H}}} \succeq \bf{0},
\end{aligned}
\end{equation}
by denoting the constellation alphabet by polynomial constraints \cite{Ma_tutorial}: $(a-p_M)...(a-p_1)(a-p_2)(a+p_1)(a+p_2)...(a+p_M)$, where $p_1, p_2, ...,$ and $p_M$ are constellation points. For the purpose of elucidation, the polynomial expression for a $4$-PAM modulation becomes: $(a-3)(a-1)(a+1)(a+3) = b^2 - 10b +1$. Using $a^{2} = b$ the problem is detailed as
\begin{equation}
\begin{aligned}
\min_{\textbf{X}} \quad & \bf{ tr } \{ \bf{ \Omega X}\} \\
\textrm{s.t.} \quad & \text{diag}
\{\bf{X}_{\text{1}...\textit{N},\text{1}...\textit{N}}\} - \bf{X}_{\textit{N}+\text{1}...\text{2}\textit{N},\text{2}\textit{N}+\text{1}} = 0\\
& \text{diag}
\{\bf{X}_{\textit{N}+\text{1}...\text{2}\textit{N},\textit{N}+\text{1}...\text{2}\textit{N}}\} \\
& - 10\bf{X}_{\textit{N}+\text{1}...\text{2}\textit{N},\text{2}\textit{N}+\text{1}} + 9 = 0\\
& \bf{X}_{\text{2}\textit{N}+\text{1},\text{2}\textit{N}+\text{1}} = 1,
\end{aligned}
\end{equation}
where the objective function in 
(\ref{MLSE_to_detect_FTN}) is written in the form of $\bf{ tr } \{ \bf{ \Omega X}\}$ using 

\begin{equation}
\Omega = 
\begin{bmatrix}
\bf{V^{\textit{H}}V} & \bf{0} & \bf{-V^{\textit{H}}y}\\
\bf{0} & \bf{0} & \bf{0}\\
\bf{-y^{\textit{H}}V} & \bf{0} & \bf{y^{\textit{H}}y}
\end{bmatrix}_{\text{2}\textit{N}+\text{1 x 2}\textit{N}+\text{1}}, 
\end{equation}
\\
and 
\\
\begin{equation}
\bf{X} =  
\begin{bmatrix}
\bf{a} & \bf{b} & \bf{1}
\end{bmatrix}_{\text{2}\textit{N}+\text{1 x 1}}
\begin{bmatrix}
\bf{a^{\textit{H}}} & \bf{b^{\textit{H}}} & \bf{1}
\end{bmatrix}_{\text{1 x 2}\textit{N}+\text{1}},
\end{equation}\\
as in \cite{luo2010semidefinite}.
\\
Here, the variable matrix $\bf{X}$ that includes the symbol sequence is optimized and the optimization outputs $\bf{x}$ whose elements are $x_{n}$, $n = 1,2,..,N$.
Note that the size $N$ of the polynomial constraints in the detection problem correspondingly expands as the modulation level increases. This is because the modulation alphabet expands with the modulation level, which requires more terms in the polynomial expression.

The true LLR of the $m$th bit $c^m_{n}$ of the $n$th transmit symbol $a_{n}$ is calculated as follows:
\begin{equation}\label{llr_expression}
\begin{aligned}
\lambda^m_{n} &= \ln{ \frac{Pr(c^m_{n}=0|x_n)}{Pr(c^m_{n}=1|x_n)}}\\
&=  \ln{\frac{p(x_{n} | c_{n}^{m} = 0) Pr(c_{n}^{m} = 0)/p(x_{n})}{p(x_{n} | c_{n}^{m} = 1) Pr(c_{n}^{m} = 1)/p(x_{n})}}\\
&=  \ln{\frac{p(x_{n} | c_{n}^{m} = 0)}{p(x_{n} | c_{n}^{m} = 1)}}
\end{aligned}
\end{equation}\\
where $Pr(c_{n}^{m} = 0) = Pr(c_{n}^{m} = 1) = 1/2$. Assuming $x_n \sim \mathcal{N}(\mu, \sigma^2)$ i.i.d., 
lookup tables are generated offline to get the LLRs, $\lambda^m_{n}$ directly instead of searching for all of the vector space. We sort the stages of generation of the LUTs for $M$-PAM as follows:
\begin{itemize}
\item Set $\frac{E_s}{No}$ in uncoded SDR detector based on $\frac{E_{b}}{N_o}$, modulation level, and the code rate that will be operated.
\item Transmit $R$ symbol sequences of length $N$.
 \item Pass them through the channel.
 \item Get the outputs $x_{n}$ of the uncoded SDR based FTN detector.
 \item Obtain the histogram of the $x_{n}$ by dividing the range of [min$(1-2^{m+1})$ max$(2^{m+1}-1)$], $m = 0, 1, ..., \text{log}_2M-2, \text{log}_2M-1$, in which interval the detector outputs, to $J$ bins. Based on the positions of transmit bits, $c_{n}^{m}$ of the symbol $a_n$, calculate the conditional probability density functions $p(x_{n} | c_{n}^{m} = 1)$, $p(x_{n} | c_{n}^{m} = 0)$ in (\ref{llr_expression}) using the output of the detector $x_{n}$. Please note that while obtaining the conditional pdfs we do not use the the matched filter output $y$ but the detector output $x_n$, maximum likelihood decoding of which gives the estimated symbol $\hat{a}$.
 \item Get the look-up tables by calculating the LLRs 
 for every bit of the symbol for the $\frac{E_s}{N_o}$ dB of interest.
 \item Apply the curvefitting operation to the tables to remove the discontinuities in them that causes the channel decoder not to work well. In the curvefitting stage, we break the unfitted curve in Figs. \ref{lamda0_tau08} and \ref{lamda1_tau08} into three parts. In the first and third parts of the curves corresponding to the ranges of $(-3,-1)$ and $(1, 3)$, respectively, we fit them using the first-degree polynomial. When it comes to the range of $(-1, 1)$ the second- and third-degree polynomials are fitted with the mentioned part for $\lambda_n^0$ and $\lambda_n^1$, respectively.
 \item Find the LLR index corresponding to the detector output using 
\begin{equation}
\begin{aligned}
    x^{n}_{index} &= \text{min}(\text{max}(\text{floor}((x_{n}\\
    &+\text{max}(2^{m+1}-1))J/M)+1, 1), J),
\end{aligned}
\end{equation} 
and pick the corresponding LLR $\lambda_{n}^{m}$ from LUT.
\end{itemize}
\section{Complexity Analysis}{\label{sec:complexity_sdr}}
\vspace{6pt}
Since the number of iterations by the numerical solver is polynomial in size of the problem (the number of variables and constraints) as expressed in \cite{luo2010semidefinite} the semidefinite relaxation based optimization method is polynomial time complexity, like $\mathcal{O}(N^{3})$. 
In the complexity comparison, we do not take the offline generation of the LUTs into account, thus, the soSDR can be said to have a similar complexity to the hard-output SDR detector. While the soSDR has still polynomial time complexity for higher modulation levels it has a significant advantage over the BCJR detector as the soSDR's complexity does not depend on the ISI length with which the BCJR detector has an exponentially increasing complexity\cite{anderson2013faster}.  

\section{Simulation Results}\label{sec:simulation}
\vspace{6pt}
In this section, we evaluate the performance of the proposed soSDR detector for a coded FTN signaling system over an AWGN channel. The proposed soSDR is simulated for the value of $\tau = 0.8$. In the simulations, we work on the $4$-PAM with Gray coding and a root raised cosine filter with roll-off factor $\alpha = 0.5$ are adopted. In the generation of the look-up tables, we set the number of transmit sequences $R$ to $10000$, the number of bins $J$ to $300$; hence, $\Delta t = 6/300 = 0.02$ in the pdf figures. For channel coding, polar codes of rate of $3/4$ and codeword length of $N = 256$ are used.
\begin{figure}[h]
\centering
\includegraphics[width=0.5\textwidth]{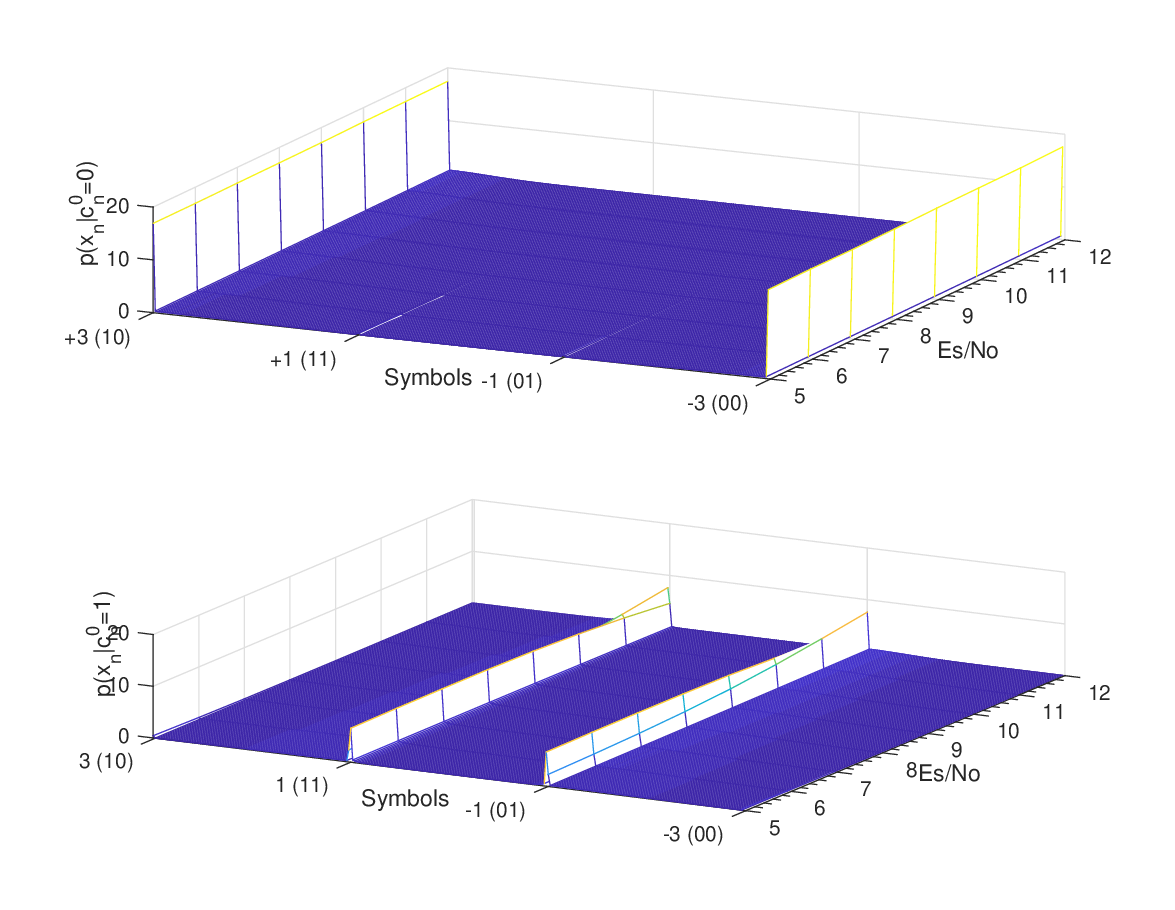}
\caption{Probability density functions of $x_n$ given $c_{n}^{0} = 0$ and $c_{n}^{0} = 1$ for the soSDR detector.}
\label{fxc0_0_and_1_tau08}
\end{figure}

\begin{figure}[b]
\centering
\includegraphics[width=0.5\textwidth]{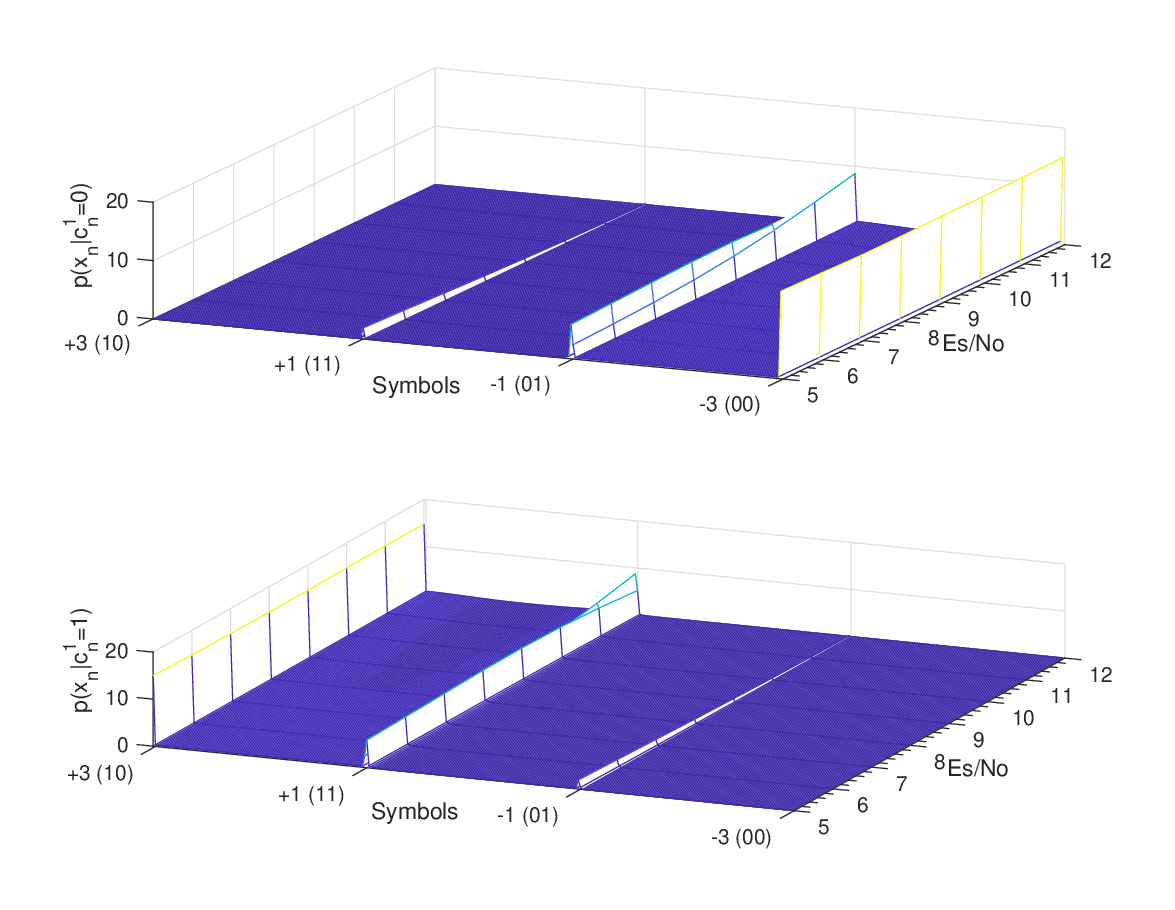}
\caption{Probability density functions of $x_n$ given $c_{n}^{1} = 0$ and $c_{n}^{1} = 1$ for the soSDR detector.}
\label{fxc1_0_and_1_tau08}
\end{figure}

Figs \ref{fxc0_0_and_1_tau08} and \ref{fxc1_0_and_1_tau08} illustrate the conditional probability density functions (pdfs) $p(x_{n} | c_{n}^{0} = 0)$ and $p(x_{n} | c_{n}^{0} = 1)$ for the least significant bit (LSB), along with the conditional pdfs $p(x_{n} | c_{n}^{1} = 0)$ and $p(x_{n} | c_{n}^{1} = 1)$ for the most significant bit (MSB), respectively. When the LSB of the transmitted symbol is $0$, it is anticipated that the likelihood of the detected symbols being $-3$ or $3$ is high, given that the LSB is $0$ exclusively for those symbols, as evident in Fig. \ref{fxc0_0_and_1_tau08}.

\begin{figure}[t]
\centering
\includegraphics[width=0.5\textwidth]{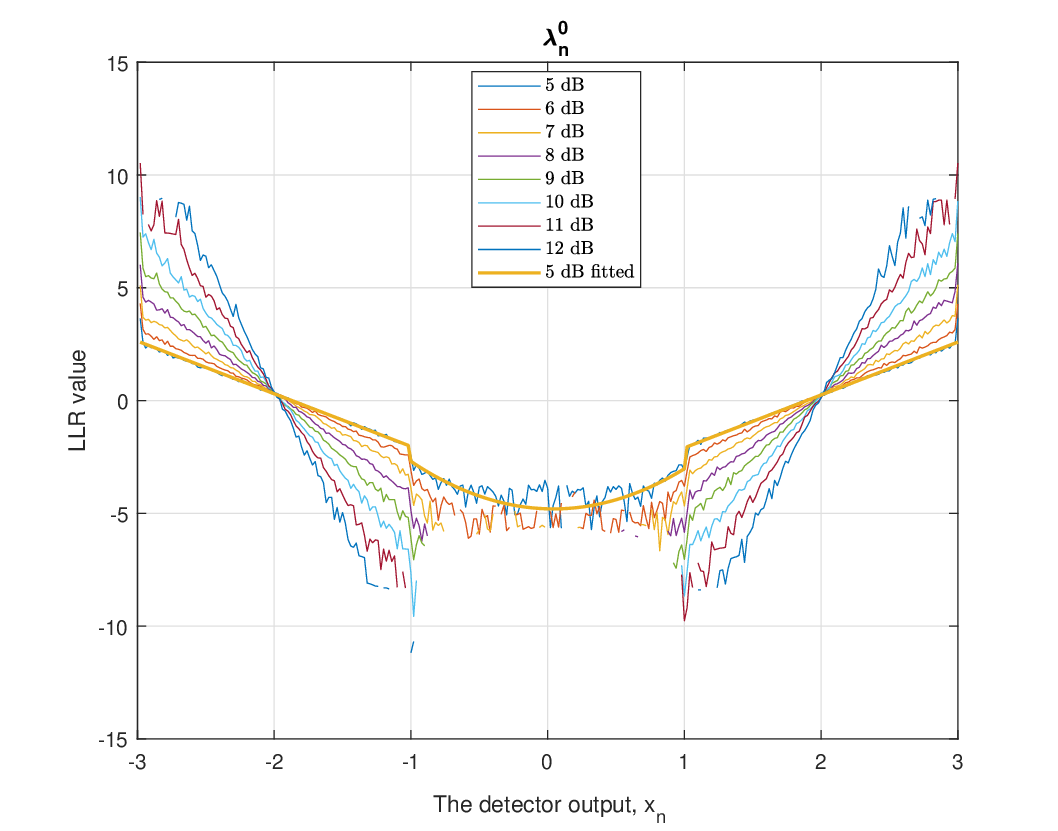}
\caption{LLR values generated for the least significant bit.}
\label{lamda0_tau08}
\end{figure}
Fig. \ref{fxc1_0_and_1_tau08} also shows that the detector may be able to incorrectly identify the symbols at lower $\frac{Es}{No}$ values, meaning less reliable LLR in the MSB position. Nevertheless, with an increase in $\frac{Es}{No}$, the received signal is more likely to fall within the correct decision regions.

\begin{figure}[h]
\centering
\includegraphics[width=0.5\textwidth]{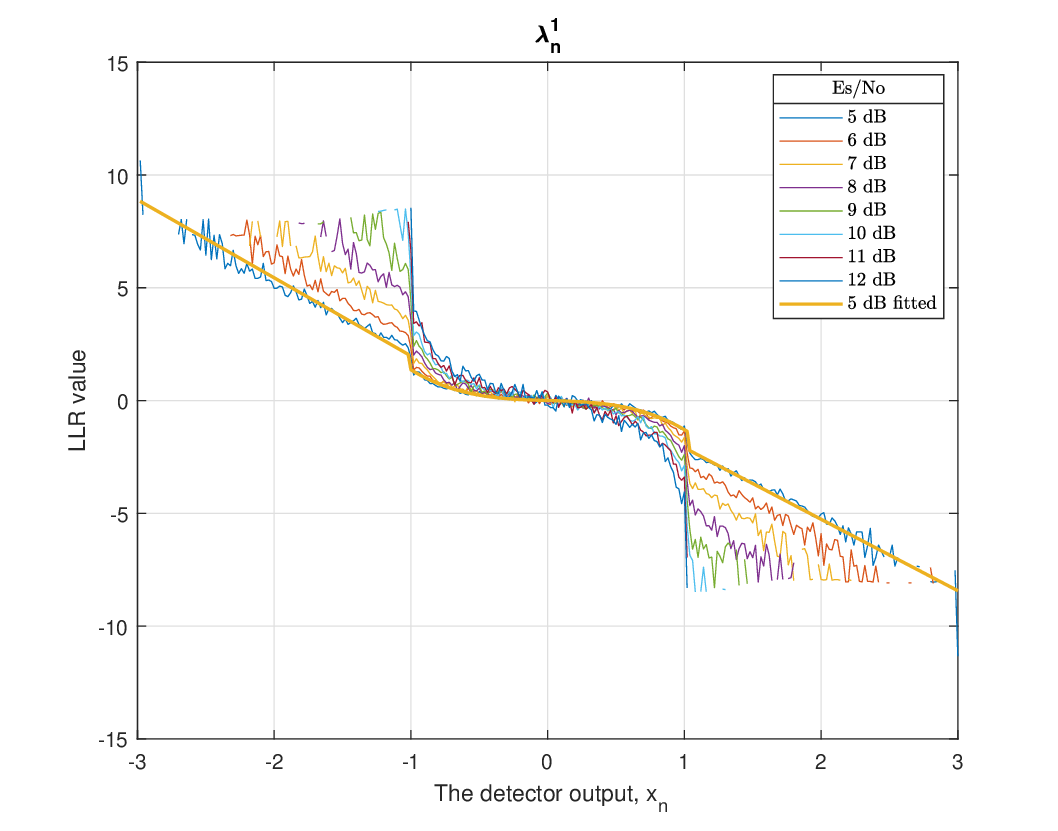}
\caption{LLR values generated for most significant bit.}
\label{lamda1_tau08}
\end{figure}

\begin{figure}[h]
\centering
\includegraphics[width=0.5\textwidth]{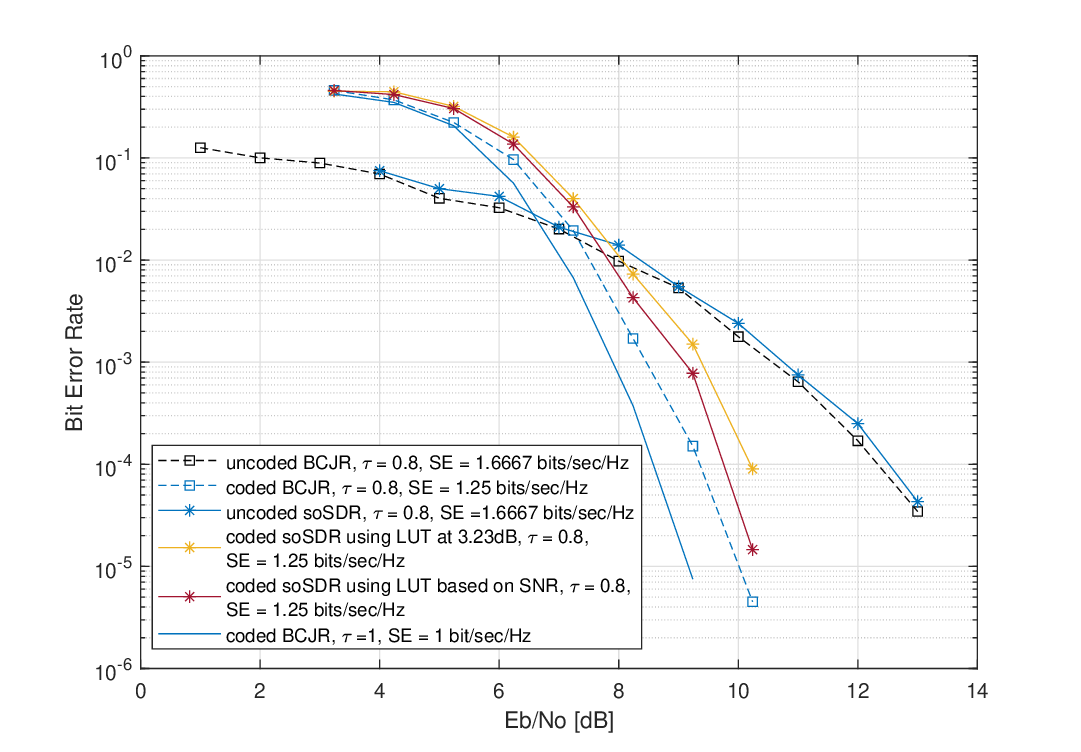}
\caption{Uncoded and coded performance of the dedector.}
\label{coded_uncoded_SDR_WMF_16QAM}
\end{figure}

Furthermore, the LLRs for the LSB (least significant bit) and MSB (most significant bit) are plotted at the different $\frac{Es}{No}$ values in Figs. \ref{lamda0_tau08} and \ref{lamda1_tau08}. As can be seen, there are some discontinues in the LLRs. This is why we use the curvefitting operation. According to the value at the output of the detector the fitted value is taken from the LSB and MSB tables. Finally, we demonstrate the BER performance of the uncoded and polar coded FTN system with the soSDR detector. At the BER of $10^{-4}$, the detector almost follows the BCJR performance, there is a gap of $0.1$ dB between them at the BER of $10^{-4}$. and it also achieves a SE of $1.6667$ bits/sec/Hz in the uncoded scenario, compared to the SE of $1.3333$ bits/sec/Hz in the Nyquist signaling, meaning that FTN signaling can provide $\frac{1.6667-1.3333}{1.6667} = 25\%$ more SE than Nyquist signaling. In the coded case, with the code rate (CR) of $\frac{3}{4}$, the SE reduces to $1.25$ bits/sec/Hz and at the same BER of $10^{-4}$, the performance of the detector is $0.8$ dB away from the BCJR, using the LUT at $\frac{E_s}{N_o}$ = $5$ dB; i.e, at $\frac{E_{b}}{N_o}$ = $3.23$ dB since $\frac{E_{b}}{N_o}$ = $\frac{E_s}{N_o}$ - $10\text{log}_{10}{(m)}$ - $10\text{log}_{10}{(\text{CR})}$. Moreover, if the lookup table based on each signal-to-noise ratio (SNR) per symbol is used the detector can approach $0.3$ dB more to the BCJR, showing how the SNR based lookup tables affect the performance in favor of the detector. Lastly, it can be seen that the SE gain in the FTN signaling may come at the cost of performance when compared to the Nyquist signaling.

\section{Conclusion}\label{sec:conc}
\vspace{6pt}
FTN signaling, with its increased detection complexity, emerges as a viable option for bandwidth-efficient transmission, showing potential for integration into future communication systems. To fully understand the practical benefits of the FTN signaling, it is necessary to explore low-complexity soft-output detection algorithms, especially when channel coding is involved. This highlights the need for generating reliable soft outputs.
In our research, we introduced the soSDR FTN detector, which can be extended to higher modulation levels. Notably, the added complexity for this detector is minimal, achieved mainly through the offline creation of LUTs based on the detector's outputs. We evaluated its performance through simulations, comparing it to the well-established BCJR, recognized for its reliable LLRs. In the uncoded case, our detector was only $0.1$ dB behind the BCJR at a BER of $10^{−4}$. In a coded setting, the difference between the soSDR detector and the BCJR was $0.8$ dB at the same BER, but this gap narrowed to $0.5$ dB when using SNR-based LUTs. With this approach, The soSDR FTN detector provides a sensible equilibrium between performance and complexity in relation to the produced LLRs.


\bibliographystyle{IEEEtran}
\bibliography{IEEEabrv,IEEE_ICC_v4.bbl}

\begin{thebibliography}{10}
\providecommand{\url}[1]{#1}
\csname url@samestyle\endcsname
\providecommand{\newblock}{\relax}
\providecommand{\bibinfo}[2]{#2}
\providecommand{\BIBentrySTDinterwordspacing}{\spaceskip=0pt\relax}
\providecommand{\BIBentryALTinterwordstretchfactor}{4}
\providecommand{\BIBentryALTinterwordspacing}{\spaceskip=\fontdimen2\font plus
\BIBentryALTinterwordstretchfactor\fontdimen3\font minus
  \fontdimen4\font\relax}
\providecommand{\BIBforeignlanguage}[2]{{%
\expandafter\ifx\csname l@#1\endcsname\relax
\typeout{** WARNING: IEEEtran.bst: No hyphenation pattern has been}%
\typeout{** loaded for the language `#1'. Using the pattern for}%
\typeout{** the default language instead.}%
\else
\language=\csname l@#1\endcsname
\fi
#2}}
\providecommand{\BIBdecl}{\relax}
\BIBdecl

\bibitem{anderson2013faster}
J.~B. Anderson, F.~Rusek, and V.~Öwall, ``Faster-than-{N}yquist signaling,''
  \emph{Proceedings of the IEEE}, vol. 101, no.~8, pp. 1817--1830, Aug. 2013.

\bibitem{modenini2014faster}
A.~Modenini, F.~Rusek, and G.~Colavolpe, ``Faster-than-{Nyquist} signaling for
  next generation communication architectures,'' in \emph{Proc. European Signal
  Processing Conference (EUSIPCO)}, Lisbon, Portugal, Sep. 2014, pp.
  1856--1860.

\bibitem{ZouOpticFTN}
D.~Zou, F.~Li, W.~Wang, W.~Ni, Q.~Sui, X.~Yi, C.~Lu, and Z.~Li, ``Simplified
  {THP} and {M}-{L}og-{MAP} decoder based faster than {N}yquist signaling for
  intra-datacenter interconnect,'' \emph{Journal of Lightwave Technology},
  vol.~41, no.~19, pp. 6300--6309, 2023.

\bibitem{CicekCIFTN}
A.~Cicek, E.~Cavus, E.~Bedeer, I.~Marsland, and H.~Yanikomeroglu, ``Coordinate
  interleaved faster-than-{N}yquist signaling,'' \emph{IEEE Communications
  Letters}, vol.~27, no.~1, pp. 229--233, Jan. 2023.

\bibitem{David8805289}
K.~David, J.~Elmirghani, H.~Haas, and X.-H. You, ``Defining 6{G}: Challenges
  and opportunities [from the guest editors],'' \emph{IEEE Vehicular Technology
  Magazine}, vol.~14, no.~3, pp. 14--16, 2019.

\bibitem{Sina_DL_based_LSD}
S.~Abbasi and E.~Bedeer, ``Deep learning-based list sphere decoding for
  faster-than-{N}yquist ({FTN}) signaling detection,'' in \emph{2022 IEEE 95th
  Vehicular Technology Conference: (VTC2022-Spring)}, Jun. 2022, pp. 1--6.

\bibitem{Sina_classification_approach}
------, ``Low complexity classification approach for faster-than-{N}yquist
  ({FTN}) signaling detection,'' \emph{IEEE Communications Letters}, vol.~27,
  no.~3, pp. 876--880, Mar. 2023.

\bibitem{bedeer2017low}
E.~Bedeer, M.~H. Ahmed, and H.~Yanikomeroglu, ``Low-complexity detection of
  high-order {QAM} faster-than-{Nyquist} signaling,'' \emph{IEEE Access},
  vol.~5, pp. 14\,579--14\,588, 2017.

\bibitem{bedeer2019low}
E.~Bedeer, H.~Yanikomeroglu, and M.~H. Ahmed, ``Low complexity detection of
  {$M$}-ary {PSK} faster-than-{N}yquist signaling,'' in \emph{Proc. IEEE
  Wireless Communications and Networking Conference (WCNC) - Workshop on
  Ultra-High Speed, Low Latency and Massive Connectivity Communication (UHSLLC)
  for 5G/B5G}, Marrakech, Morocco, May 2019, pp. 1--6.

\bibitem{PrljaWMF}
A.~Prlja, J.~B. Anderson, and F.~Rusek, ``Receivers for faster-than-nyquist
  signaling with and without turbo equalization,'' in \emph{IEEE International
  Symposium on Information Theory}, July 2008, pp. 464--468.

\bibitem{leroux2011hardware}
C.~{Leroux}, I.~{Tal}, A.~{Vardy}, and W.~J. {Gross}, ``Hardware architectures
  for successive cancellation decoding of polar codes,'' in \emph{Proc. IEEE
  International Conference on Acoustics, Speech and Signal Processing
  (ICASSP)}, Prague, Czech Republic, May 2011, pp. 1665--1668.

\bibitem{Ma_tutorial}
W.-K. Ma, \emph{Semidefinite Relaxation and Its Application in Signal
  Processing and Communications}.\hskip 1em plus 0.5em minus 0.4em\relax Xi'an,
  China: MIIS Tutorial, Jul. 2012.

\bibitem{luo2010semidefinite}
Z.-Q. Luo, W.-K. Ma, A.~M.-C. So, Y.~Ye, and S.~Zhang, ``Semidefinite
  relaxation of quadratic optimization problems,'' \emph{{IEEE} Signal Process.
  Mag.}, vol.~27, no.~3, pp. 20--34, May 2010.

\end{thebibliography}
\end{document}